\documentclass[12pt,preprint]{aastex}
\usepackage{latexsym,graphicx,natbib}

\newcommand{\beq}{\begin{equation}}
\newcommand{\eeq}{\end{equation}}

\begin{document}
\bibliographystyle{apj}

\title{Photometry of Irregular Satellites of Uranus and Neptune}

\shorttitle{Photometry of Irregular Satellite}
\shortauthors{Grav, Holman \& Fraser}
\medskip

\author{Tommy~Grav\altaffilmark{1}}
\affil{Harvard-Smithsonian Center for Astrophysics, MS51, 60 Garden Street, Cambridge MA 02138}
\email{tgrav@cfa.harvard.edu}
\author{Matthew~J.~Holman\altaffilmark{2}}
\affil{\footnotesize \it Harvard-Smithsonian Center for Astrophysics, MS51, 60 Garden Street, Cambridge, MA 02138}
\email{mholman@cfa.harvard.edu}
\medskip
\author{Wesley C. Fraser\altaffilmark{2}}
\affil{\footnotesize \it McMaster University, Hamilton, ON L8S 4M1, Canada}
\email{fraser@physics.mcmaster.ca}
\altaffiltext{1}{Visiting Astronomer, Keck Observatory}
\altaffiltext{2}{Visiting Astronomer, Magellan Observatory}
\date{\rule{0mm}{0mm}}

%-------------------------------------------
\begin{abstract}
We present BVR photometric colors of six Uranian and two Neptunian
irregular satellites, collected using the Magellan Observatory (Las
Campanas, Chile) and the Keck Observatory, (Manua Kea, Hawaii). The
colors range from neutral to light red, and like the Jovian and the
Saturnian irregulars \citep{Grav.2003b} there is an apparent lack of
the extremely red objects found among the Centaurs and Kuiper belt
objects. 

The Uranian irregulars can be divided into three possible dynamical
families, but the colors collected show that two of these dynamical
families, the Caliban and Sycorax-clusters, have heterogeneous
colors. Of the third possible family, the 168-degree cluster
containing two objects with similar average inclinations but quite
different average semi-major axis, only one object (U~XXI~Trinculo)
was observed.  The heterogeneous colors and the large dispersion of
the average orbital elements leads us to doubt that they are
collisional families.  We favor single captures as a more likely
scenario. The two neptunians observed (N~II~Nereid and S/2002~N1) both
have very similar neutral, sun-like colors. Together with the high
collisional probability between these two objects over the age of the
solar system~\citep{Nesvorny.2003a,Holman.2004}, 
this suggests that S/2002~N1 be a fragment of Nereid, broken loose
during a collision or cratering event with an undetermined impactor.

\end{abstract}
%-------------------------------------------
\section{Introduction}

Irregular satellites are small bodies that orbit their parent planets
in large, eccentric orbits with high inclinations relative to the
planet's equatorial plane. Irregular satellites have been discovered
around all the giant planets and are thought to have been captured
from heliocentric orbits during the last stages of the formation of the
giant planets.  While objects in heliocentric orbits may be
temporarily captured in planetocentric orbits, some loss of energy is
needed to make the capture permanent. Several processes have been
proposed for this change of orbital energy: 1) an increase in the mass of the
planet through accretion \citep{Heppenheimer.1977}; 2) gas drag
through an extended envelope or disk around the still forming planet
\citep{Pollack.1979}; 3) collision or close encounters with a
pre-existing regular moon or another temporarily captured object
\citep{Colombo.1971}; 4) dynamical friction from a large number of
small outer solar system bodies \citep{Astakhov.2003, Goldreich.2002}.  

It was recognized by \citet{Colombo.1971}, and later by
\citet{Gladman.2001}, that the irregular satellites cluster
in groups with similar dynamical properties. This clustering is
considered to be evidence that the members of a cluster are remnants
of a larger progenitor that was captured and subsequently broken
up. In \citet{Grav.2003b} we reported optical BVRI photometry of a
large number of irregular satellites of Jupiter and Saturn, showing
that almost all of the known dynamical clusters have homogeneous colors,
supporting the thesis suggesting the fragmentation of larger
progenitors.  Determining the near-infrared colors of the brightest
Jovian and Saturnian irregular satellites revealed that the normalized
broadband reflectance spectra are very similar to the spectra of C-
and D-type outer main belt asteroids \citep{Grav.2004}.   

We now turn to Uranus and Neptune. The physical knowledge of the
irregular satellites of the two outer planets is extremely
limited. Only two Uranian and one Neptunian irregular satellites have
been the targets of photometrical studies and have had their colors
determined. \cite{Maris.2001} observed the two Uranian satellites,
U~XVI~Caliban and U~XVII~Sycorax, in BVRI filters, using the 3.6 m ESO
NTT, on La Silla, Chile. They found that both have moderately red
colors, with Sycorax being slightly bluer than Caliban. They compared
their observations to other families of objects in the solar system
and found that the two satellites were clearly redder than Uranus and
its regular satellites. They also compared the $V-R$ colors with the
histogram found in \citet{Jewitt.1996} and placed the satellites among
the bluest Kuiper belt objects and the reddest near-earth objects.  

\citet{Romon.2001} performed a more detailed study of Sycorax, also
determining the BVRI colors using the 3.5m Telescopio Nazionale
Galileo, La Palma. They, however, added photometric observations  in
the J-band and spectroscopic observations in the near-infrared using
ESO's 8m Very Large Telescope in Chile. They compared their results
with that of other small bodies in the Solar System, and argued that
Sycorax is more similar to TNOs, Centaurs and cometary nuclei,
than to the Trojans and irregular satellites of Jupiter.  

N~II~Nereid was discovered in 1949 by G. Kuiper. Due to its brightness
it has been extensively studied. Colors have been determined by
\citet{Schaefer.2000}, and \citet{Brown.1999} used near-infrared
spectra to show that the satellite has features indicative of water
ice. Voyager II observations during its fly-by of Neptune were
used to determine the albedo, $p\sim0.2$, of Nereid
\citep{Thomas.1991}. It is thus believed that Nereid is an icy
body.  It is either an inner satellite gravitationally scattered as
N~I~Triton was captured or a captured ice body originating in the
Kuiper belt~\citep{McKinnon.1984,Goldreich.1989}.

In this paper we report the BVR colors of six Uranian and two Neptunian
irregular satellites, and discuss the implications of the colors
determined. In section \ref{sec:obs} we describe the observations
performed and our method of data reduction. Section \ref{sec:results}
contains the results and interpretations of the observations. 

\section{The Observations}
\label{sec:obs}

The observations presented here were performed at the 6.5m Clay
telescope at the Magellan Observatory using the MagIC instrument and
at the 10m Keck II telescope using the DEIMOS instrument.  The
observations at the Clay telescope were performed on July
27th and 28th, 2003. The conditions were photometric and observations
of Caliban, Sycorax, U~XX~Stephano, Nereid and S/2002~N1 were
performed . The observations using the Keck were done on August 1,
2003. The first part of the night was subject to high cirrus clouds,
that moved off to the horizon at about midnight. The remainder of the
night was photometric and used to observe the uranian satellites
U~XVIII~Prospero, U~XIX~Setebos and U~XXI~Trinculo, as well as the
neptunian satellite S/2002~N1. 

A number of standard stars \citep{Landolt.1992} were observed each night,
covering the same elevation as the science targets. Transformation
equations containing zero points, airmass corrections and color
corrections were determined and used to determine the V-magnitude and colors
(B-V and V-R) of the science targets. The DAOPHOT package under the
IRAF environment was used for the all the data reduction.  
  
\subsection{Magellan MagIC}

The MagIC instrument is a single SITe 2048x2048 CCD camera with a
rather small field of view, 2 by 2 arcminutes. For the MagIC
observations a Harris BVR filter set was used. The seeing during the
observations varied from $1.2-2.0$ arcseconds. We performed aperture
photometry using a inner and outer aperture of $1.725$ and $4.14$
arcseconds, respectively. The aperture correction were small
($0.05-0.21$ magnitudes) in all filters. Using the observed Landolt
standars stars we determined the following transformation equations: 
\begin{eqnarray}
	B &=& b + 26.80 - 0.22 a \\ 
	V &=& v + 26.94 - 0.21 a  \\
	R &=& r + 27.15 - 0.15 a
\end{eqnarray}
where $b$, $v$ and $r$ are the instrumental magnitudes and $a$ is the
airmass. Color corrections were included in the determination of the
transformation equations, but were negligible.  

\subsection{Keck II DEIMOS}

The DEep Imaging Multi-Object Spectrograph (DEIMOS) is an optical
wavelength imaging spectrograph. We used the instrument in direct
imaging mode. The seeing varied from $0.8-1.2$ arc-seconds. We again
used aperture photometry, using an inner and outer aperture of $0.95$
and  $2.37$ arc-seconds, respectively. The aperture corrections were
also small ($0.05-0.29$ magnitudes) in all filters. 

The filters available for the DEIMOS are rather special and do not
conform to any of the usual filter systems. We therefore included
color corrections in the determination of the transformation equations
from instrumental to photometric magnitude. The color corrections in
the B and V filters were non-negligible but still small given the
moderate colors of our targets. For the Keck II observations we
derived the following transformation equations: 
\begin{eqnarray}
	B &=& b + 27.33 - 0.24 a + 0.23 (B - V) \\ 
	V &=& v + 27.90 - 0.24 a - 0.15 (B - V) \\
	R &=& r + 28.06 - 0.15 a
\end{eqnarray}
where, again, $b$, $v$ and $r$ are the instrumental magnitudes and $a$
is the airmass.  

\section{The Results}
\label{sec:results}

\begin{figure}[htbp]
	\begin{center}
		\includegraphics[width=8cm,height=8cm]{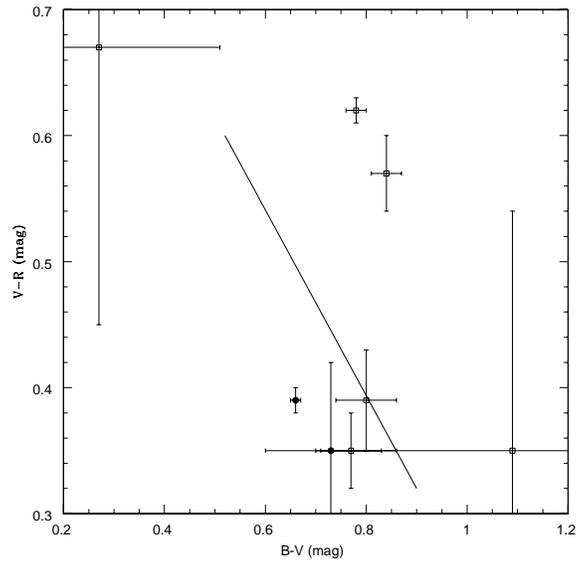}
		\caption{Grav, Holman \& Fraser (2004)}	
   		\label{fig:bvr}
  	\end{center}
\end{figure}

The data collected is presented in Table \ref{tab:res} and plotted in
a standard B-V vs V-R diagram in Figure \ref{fig:bvr}.  The colors are
similar to those found among the Jovian and Saturnian, and the colors
seem to be separated into two groups. One is essentially {\bf neutrally
colored} containing Prospero, Setebos, the two Neptunians observed,
and possibly Trinculo. The other is {\bf slighty red} and contains the
two large Uranians, Caliban and Sycorax. The Uranian irregular
Stephano may, due to its large error bars, be put in either of the
two groups. Interestingly, neither the Uranian or the Neptunian
irregular satellites have members with the extremely red colors found
among the Kuiper belt objects.  

Caliban, Sycorax, and Nereid have had their colors determined
previously \citep{Schaefer.2000, Maris.2001, Romon.2001}. The colors
determined in this paper is in excellent agreement with the results of
these papers, except for the B-V colors of Caliban and Sycorax
determined by \citet{Maris.2001}. Their B-V colors are higher than and
inconsistent with ours even at the $3\sigma$ level.  We are unable to
explain this discrepancy.    

Using the geometric circumstances of the observations (see Table
\ref{tab:obs}) we have used the observed V-magnitude to derive
absolute magnitudes at zero phase angle and unit heliocentric and
geocentric distances.  
The equation used is
\begin{equation}
	m_V(1,1,0) = V - 5 \log(\Delta r) - \beta \alpha
\end{equation}
where V is the observed magnitude, $\alpha$ is the phase angle,
$\beta$ is the phase angle correction factor, and $\Delta$ and $r$ are
the geocentric and heliocentric distances, respectively. While the
distances and the phase angle are well known quantities, the linear phase
coefficient, $\beta$, is highly uncertain. \citet{Schaefer.2001} observed 
Nereid at a range of phase angles and and found $\beta = 0.38$ for 
$\alpha < 1^\circ$. This value is similar to the phase coefficients of small, 
inner Uranian satellites \citep{Karkoschka.2001}. We will use this value 
to estimate the absolute magnitudes for both the Uranian and Neptunian 
irregulars. We use the derived absolute magnitudes to determine the 
sizes of the satellites, by applying the equation: 
\begin{equation}
	D = \frac{1329 \cdot 10^{[-m_V/5]}}{\sqrt{p}}
\end{equation}
where $p$ is the geometric albedo of the satellite in the visual. The
geometric albedo of the irregular satellites is another highly unknown
property. Of the Uranian and Neptunian only Nereid has had its albedo
determined at $p=0.2$ \citep{Thomas.1991}. This value is significantly
larger than that of the small inner neptunians ($p=0.06$), the small
inner Uranian satellites ($p=0.07$), J~VI~Himalia ($p=0.05$) and
S~IX~Phoebe ($p=0.08$). We have based on this chosen to use $p=0.07$
for the Uranian irregulars and $p=0.2$ for the Neptunian
irregulars. The resulting sizes are given in Table \ref{tab:obs}.

\subsection{The Uranian Irregular Satellites}

\begin{figure}[htbp]
  \begin{center}
  \includegraphics[width=8cm,height=8cm]{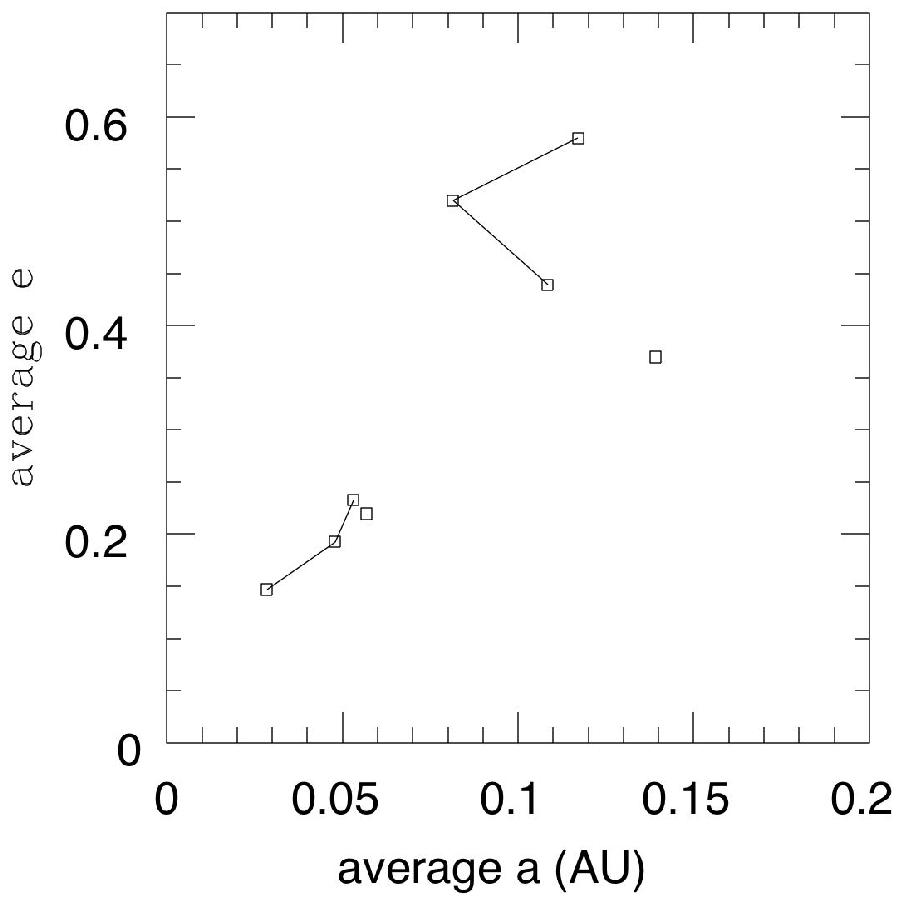}
		\caption{Grav, Holman \& Fraser (2004)}
  \label{fig:uranus_ae}
  \end{center}
\end{figure}

\begin{figure}[htbp]
  \begin{center}
  \includegraphics[width=8cm,height=8cm]{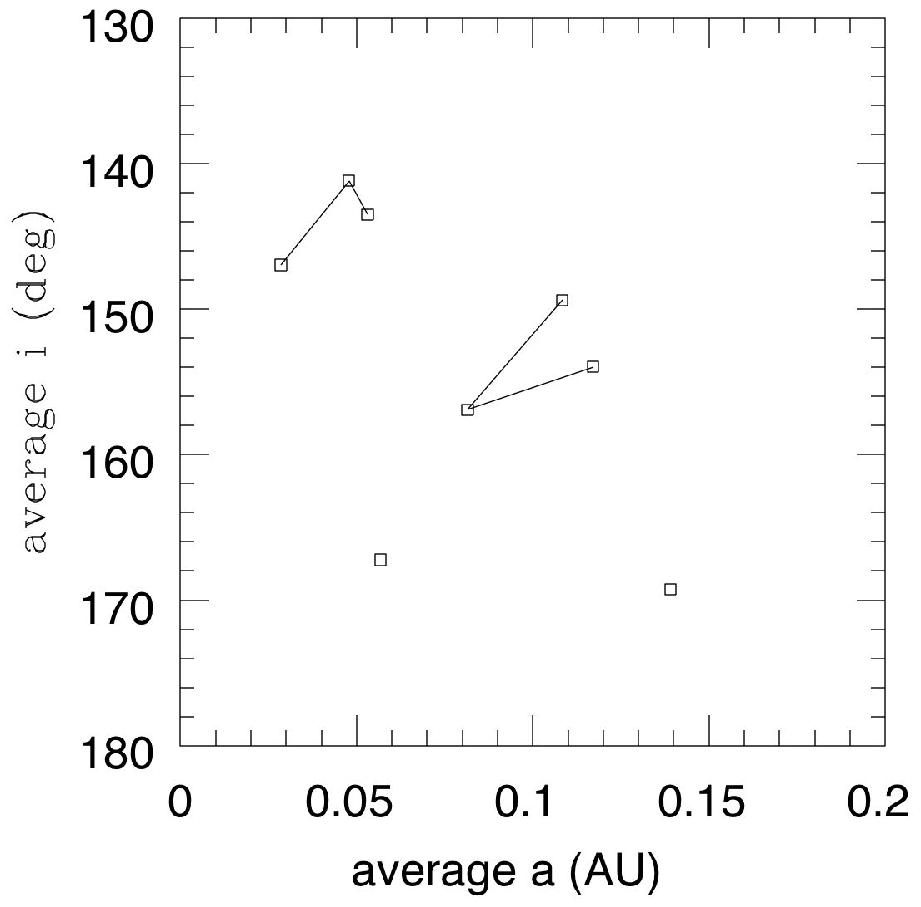}
		\caption{Grav, Holman \& Fraser (2004)}
  \label{fig:uranus_ai}
  \end{center}
\end{figure}

Unfortunately, the colors of the Uranian irregular satellites offer
little information about the satellites' origin.  To check for
dynamical families among the Uranian irregulars, we performed
long-term ($t \sim 10^8$years) integrations of the nominal orbits of
the known irregular satellites, and our resulting average elements are
similar to those found by \cite{Nesvorny.2003a} (the average elements
found in our integrations are plotted in Figure \ref{fig:uranus_ae}
and \ref{fig:uranus_ai}, which includes three Uranian irregulars not
found in \citet{Nesvorny.2003a}). 

From the average orbital elements we divide the known Uranian
irregulars into two possible dynamical families. The Caliban-family
include Caliban, Stephano and S/2001~U3, while the Sycorax-family
consists of Sycorax, Prospero and Setebos. This leaves Trinculo,
S/2001~U1 and S/2003~U3 (the only prograde uranian irregular satellite
known to date) as single objects. It is interesting to note that
Trinculo and S/2001~U1 have very similar inclinations, similar to the
two prograde Saturnian irregular satellites, and could thus be a
possible third dynamical family or perhaps the results of a different
capturing process than the other Uranian irregulars. Their separation
in average semi-major axis is significantly larger, $\Delta a =
0.08$AU, than that of the Saturnian Inuit and Gallic clusters ($\Delta
a_I = 0.01$AU and $\Delta a_G = 0.04$AU, respectively). 

It is thus clear that if the known Uranian irregular satellites are
indeed clustered into dynamical families, these two families have
heterogeneous colors. It should be noted that this is not necessarily
hard evidence against family structure. The Hilda asteroids, for
example, have an apparent spectral slope-size, with the large members
being P-types and the smaller being D-type asteroids
\citep{Dahlgren.1995, Dahlgren.1997}. It is possible that the Uranian
irregular satellites have a similar size-color correlation, but the
low number of  known objects available make this theory a pure
speculation.

\subsection{Nereid and S/2002~N1}

The observed colors of Nereid and S/2002~N1 are basically the same,
suggesting similar surface compositions. Studies of the collisional 
probabilities between the irregular satellites of Neptune show that Nereid
and S/2002~N1 has a high probability of colliding ($~0.41$) over $4.5$ Gy
\citep{Holman.2004}. The similar colors of these two objects thus point to 
the possibility that S/2002~N1 is a fragment of Nereid. The nearly
spherical shape of Nereid \citep{Grav.2003a} offers no photometric evidence that
Nereid was ever catastrophically disrupted, however Voyager II images
suggest a cratered body \citep{Thomas.1991}. We suggest that S/2002~N1 is therefore ejecta 
from cratering event on Nereid. This hypothesis would be 
further supported if smaller irregular satellites with similar colors
and high collisional probabilities with Nereid were found.  The neutral colors of Nereid and 
S/2002~N1 does not give any new hint to the possible origins of the two 
satellites. The colors, albedo and spectra of Nereid are similar to both regular satellites such as Oberon and Umbriel \citep{Buratti.1991,Brown.1999}, as well as several Kuiper belt objects \citep{Tegler.2003b}. 
  
The large difference in magnitude between the two observations of
S/2002~N1 indicates that the object has a significant rotational light
curve, most likely due to an out-of-round shape. It is also possible
that the object has a significant opposition surge in its phase angle
light curve. Nereid has a significant opposition
surge~\citep{Schaefer.2001}, with an increase of $0.38$ magnitudes per 
degree. If the magnitude difference of S/2002~N1 found in this paper
is solely due to an opposition effect it implies an increase of $3.9$
magnitudes per degree, a full order of magnitude larger than that of
Nereid. This seems extremely unlikely, leading us to believe that the
magnitude difference is fully or partially due to a rotational light
curve. 

\section{Conclusions}
\label{sec:conclusions}

We have reported $BVR$ colors of six Uranian and two Neptunian
irregular satellites. The colors are similar to that of the Jovian and
Saturnian irregular satellites \citep{Grav.2003b}. It seems however
unlikely that the irregular satellites of Uranus and Neptune have
origins in the outer main asteroidal belt. The lack of extremely red
objects among the observed Uranian and Neptunian irregulars makes
their color distribution different from that of the Centaurs and
Kuiper belt, thus hinting at an origin among the outer planets.  
Note that the lack of extremely red objects could just be due to the small 
sample of objects studied. Further study of the irregular satellite population
should settle this issue.
 
The heterogeneity of the possible dynamical families in the uranian
irregular system leads us to question the veracity of the apparent
dynamical families. With both Uranus and Neptune possibly having undergone
special events (the Great Collision \citep{Brunini.2002} and the
capture of Triton \citep{Farinella.1980}), it seems uncertain whether
one would expect any dynamical families of irregular satellites to
exist around the two planets. If one chooses to invoke the hypothesis
that the Caliban- and Sycorax-clusters are indeed real families, one
has to favor the theory that the similar inclinations of Trinculo and
S/2001~U1 are due to a different capturing process than the rest of
the uranian irregulars, due to the very large separation of semi-major
axis of the two objects. Further study of this is necessary. 

\section{Acknowledgments}

Some of the data presented herein were obtained at the W.M. Keck
Observatory, which is operated as a scientific partnership among the
California Institute of Technology, the University of California and
the National Aeronautics and  Space Administration. The Observatory
was made possible by the generous financial support of the W.M. Keck
Foundation.  

Tommy Grav is a Smithsonian Astrophysical Observatory Pre-doctoral
Fellow at the Harvard-Smithsonian Center for Astrophysics, Cambridge,
USA. He is a graduate student at the Institute for Theoretical
Astrophysics at the University in Oslo, Norway. 

 This work was supported by NASA grants JPL1257944, NAG5-9678 and NAG5-10438.

%---------------------------------------------------------------
\section{Figure and Table Captions}

{\bf Figure 1.}\\
B-V vs. V-R color. The error bars show the 1-$\sigma$ errors. The dashed line is the the same line used to arbitrarily separate the Jovian and Saturnian irregular satellites into the {\it gray} and {\it light red} in \cite{Grav.2003b}.

 {\bf Figure 2.} \\
The average semi-major axis ($a$) and eccentricities ($e$) of the irregular satellites of Uranus from $10^8$ year integrations. The lines indicate possible irregular satellite families

{\bf Figure 3.} \\
The average semi-major axis ($a$) and inclinations ($i$) of the irregular satellites of Uranus from $10^8$ year integrations. The lines indicate possible irregular satellite families.

{\bf Table 1.} \\
The results of our $BVR$ color survey of the Uranian and Neptunian. The table gives the $1\sigma$ errors. The first line for S/2002~N1 gives the weighted mean values of the two observations performed.

{\bf Table 2.} \\
The geometry of the observations. The table gives the heliocentric distance, $r$, geocentric distance, $\Delta$ and phase angle, $\alpha$. The table also shows the derived absolute magnitude at opposition, unit heliocentric and geocentric distance, as well as the derived sizes.

\begin{table}[htbp]
	\begin{center}
		\begin{tabular}{lccc}
 		Object & V & B-V & V-R \\
		\hline
		U~XVI~Caliban      & $22.58 \pm 0.02$ & $0.84\pm0.03$ & $0.57\pm0.03$ \\
		U~XVII~Sycorax     & $20.94 \pm 0.01$ & $0.78\pm0.02$ & $0.62\pm0.01$ \\
		U~XVIII~Prospero  & $23.91\pm 0.03$ & $0.80\pm0.06$ & $0.39\pm0.04$ \\
		U~XIX~Setebos      & $23.88\pm 0.03$ & $0.77\pm0.06$ & $0.35\pm0.03$ \\
		U~XX~Stephano    & $25.12 \pm 0.17$ & $0.27\pm 0.24$ & $0.67\pm0.22$ \\
		U~XXI~Trinculo      & $25.25 \pm 0.18$ & $1.09 \pm 0.40$  & $0.35 \pm 0.19$ \\
\\
		N~II~Nereid            & $19.25\pm0.01$ & $0.66\pm0.01$ & $0.39\pm0.01$ \\
		S/2002 N1              &                                 & $0.73\pm0.13$ & $0.35\pm0.07$ \\
                   		               & $23.82\pm0.06$ & $0.87\pm0.10$ & $0.29\pm0.08$ \\
                                  			& $24.48 \pm 0.09$ & $0.51 \pm 0.16$ & $0.47 \pm 0.12$ 

		\end{tabular}
		\caption{Grav, Holman \& Fraser (2004)}
   		\label{tab:res} 

	\end{center}
\end{table}

\begin{table}[htbp]
   	\begin{center}
      		\begin{tabular}{llccccc}
          		Object & Telescope & r & $\Delta$ & $\alpha$ & $m_V$ & D \\
                       		&                     & (AU) & (AU)  & ($\circ$) & (1,1,0) & (km) \\
           	\hline
		U~XVI~Caliban  & Magellan   & 20.02 & 19.11 & 1.33  &   $9.16 \pm 0.04$ & $\sim 74$ \\
		U~XVII~Sycorax  & Magellan  & 20.08 & 19.18 & 1.34  &   $7.50 \pm 0.02$ & $\sim 159$ \\
		U~XVIII~Prospero & Keck II     & 20.15 & 19.20 & 1.09  & $10.56 \pm 0.05$ & $\sim 39$ \\
		U~XIX~Setebos  & Keck II        & 19.96 & 19.02 & 1.10  & $10.57 \pm 0.05$ & $\sim 39$ \\
		U~XX~Stephano & Magellan  & 20.07 & 19.16 & 1.32  & $11.69 \pm 0.17$ & $\sim 23$ \\
		U~XXI~Trinculo & Keck II         & 20.07 & 19.12 & 1.08  & $11.92 \pm 0.18$ & $\sim 21$ \\
\\
		N~II~Nereid   & Magellan        & 30.07 & 29.06 & 0.26   & $4.44 \pm 0.01$ & $\sim 384$ \\
		S/2002 N1 & Magellan             & 30.12 & 29.11 & 0.25   & $9.01 \pm 0.07$ & $\sim 47$\\
                   	 	& Keck II                  & 30.12 & 29.10 & 0.08    & $9.74 \pm 0.08$ & $\sim 34$\\

		\end{tabular}
		\caption{Grav, Holman \& Fraser (2004)}
   		\label{tab:obs} 
	\end{center}
\end{table}

%---------------------------------------------------------------
\newpage
\bibliography{ms}
%---------------------------------------------------------------
\end{document}